\documentclass[aps,pra,twocolumn,letterpaper,superscriptaddress,10pt]{revtex4-1}
\usepackage{amsmath,amssymb}
\usepackage{graphicx,ulem,mathptmx}
\usepackage[pdftex,dvipsnames,usenames]{xcolor}
\usepackage[colorlinks=true,urlcolor=blue,citecolor=blue,linkcolor=blue]{hyperref}

\begin{document}
\title{Mode-independent quantum entanglement for light}

\author{Jan Sperling}
	\affiliation{Integrated Quantum Optics Group, Applied Physics, University of Paderborn, 33098 Paderborn, Germany}

\author{Armando Perez-Leija}
	\affiliation{Max-Born-Institut, Max-Born-Stra\ss{}e 2A, 12489 Berlin, Germany}
	\affiliation{Humboldt-Universit\"{a}t zu Berlin, Institut f\"{u}r Physik, AG Theoretische Optik \& Photonik, D-12489 Berlin, Germany}

\author{Kurt Busch}
	\affiliation{Max-Born-Institut, Max-Born-Stra\ss{}e 2A, 12489 Berlin, Germany}
	\affiliation{Humboldt-Universit\"{a}t zu Berlin, Institut f\"{u}r Physik, AG Theoretische Optik \& Photonik, D-12489 Berlin, Germany}

\author{Christine Silberhorn}
	\affiliation{Integrated Quantum Optics Group, Applied Physics, University of Paderborn, 33098 Paderborn, Germany}

\date{\today}

\begin{abstract}
	We address the problem of the persistence of entanglement of quantum light under mode transformations, where orthogonal modes define the parties between which quantum correlations can occur.
	Since the representation of a fixed photonic quantum state in different optical mode bases can substantially influence the entanglement properties of said state, we devise a constructive method to obtain families of states with the genuine feature of remaining entangled for any choice of mode decomposition.
	In the first step, we focus on two-photon states in a bipartite system and optimize their entanglement properties with respect to unitary mode transformations.
	Applying a necessary and sufficient entanglement witness criteria, we are then able to prove that the class of constructed states is entangled for arbitrary mode decompositions.
	Furthermore, we provide optimal bounds to the robustness of the mode-independent entanglement under general imperfections.
	In the second step, we demonstrate the power of our technique by showing how it can be straightforwardly extended to higher-order photon numbers in multipartite systems, together with providing a generally applicable and rigorous definition of mode-independent separability and inseparability for mixed states.
\end{abstract}

\maketitle

%%%%%%%%%%%%%%%%%%%%%%%%%%%%%%%%%%%%%%%%%%%%%%%%%%%%%%%%
% Introduction
%%%%%%%%%%%%%%%%%%%%%%%%%%%%%%%%%%%%%%%%%%%%%%%%%%%%%%%%
\section{Introduction}

	Quantum entanglement is a fundamental property of quantum states in composite systems, exceeding our understanding of classical, i.e., separable, correlations \cite{HHHH09}.
	For this reason, early debates on the validity of quantum mechanics used such quantum correlations to demonstrate the unique and distinct aspects of quantum theory \cite{ERP35,S35}.
	Nowadays, quantum entanglement has evolved into a key resource for performing useful quantum computation and communication protocols \cite{NC00,GT07}.
	In particular, photonic systems constitute a versatile platform which enables us to transmit quantum information between distant parties \cite{KLM01,GT07,KMNRDM07,S09}.
	However, as optical fields can be decomposed in a variety of fundamental modes, e.g., plane-wave and Hermite-Gaussian bases, the encoding of quantum information in photons is not unique.
	Still, optical systems inherently offer the advantage of being scalable because the number of modes is not limited, rendering it possible to realize complex forms of multipartite entanglement for quantum communication between multiple nodes of a network.

	For the above reasons, entanglement theory is an exciting field of research as it combines fundamental concepts with applications in upcoming technologies \cite{HHHH09,NC00}.
	However, the separability problem, i.e., deciding whether a state is separable, is also a sophisticated and computationally hard problem \cite{G03,I07}.
	Nevertheless, during the last decades, remarkable progress has been made \cite{HHHH09,GT09}.
	This includes the theoretical classification of highly complex forms of entanglement in multipartite systems \cite{HV13,LM14,SSV14}, as well as experiments which significantly widened the class of accessible states with interesting entanglement characteristics \cite{Hetal05,Yetal13,Cetal14,GSVCRTF15,GSVCRTF16,Cetal17}, where quantum-optical implementations are in many regards pioneering when it comes to realizing, witnessing, and utilizing entanglement.

	The notion of an entangled state changes with the choice of the mode decomposition of quantum light, a property which is discussed in more detail in the continuation of this work.
	This is due to the definition of separability which presupposes a given separation of degrees of freedom \cite{W89}, which are in our case the optical modes of the quantized radiation field.
	Thus, a mode transformation can alter the quantum correlation properties of light from entangled to separable, and vice versa \cite{X02,KSBK02,VS14}.
	Consequently, one has to ask oneself if there exist states which are entangled regardless of the chosen mode decomposition.
	In this paper, we provide a positive answer to this question by deriving a constructive approach to states with this very property.

	Related obstacles occur when studying entanglement in general quantum field theories, where different superposition principles apply simultaneously.
	Thus, it is cumbersome to distinguish classical interference from quantum-mechanical ones.
	In quantum optics, for example, we can superimpose modes and quantum states \cite{MW95}.
	The former transformation is a classical operation which is based on the linear structure of Maxwell's equations.
	The latter quantum superposition can lead to entanglement when involving nonlocal states defined over multiple modes.
	In fact, the confusion between classical and quantum superpositions and tensor-product structures led to the unsuitably named concept of classical entanglement \cite{S98,KB15}.
	This confusion is further reinforced when considering classical mode transformations that can result in entanglement of initially separable states by altering the initial separation of degrees of freedom.
	For instance, this can be seen from how one can generate entangled states from single-mode nonclassical states via general beam splitter transformations \cite{X02,KSBK02,VS14}.
	This effect is, for example, used when two single photons interfere in the Hong-Ou-Mandel experiment \cite{HOM87}, which effectively corresponds to a mode transformation that results in an entangled output state.
	Thus, a challenge is to find and certify stronger forms of entanglement which are independent of classical interference, i.e., not subject to mode bases changes.

	In this paper, we construct families of multiphoton states in multimode systems which exhibit quantum entanglement for arbitrary mode representations, defining the notion of mode-independent quantum entanglement (MIQE).
	Initially, in Sec. \ref{sec:Preliminaries},  we demonstrate that the definition of entanglement between optical modes is indeed dependent on the choice of the optical mode reference, leading to the observation that at least two-photon states are required to achieve our goal.
	By combining photons which are defined in nonparallel and nonorthogonal modes, we are then able to build up states with the desired inseparability properties in Sec. \ref{sec:MN2}.
	Furthermore, we optimize the occurring free parameter to determine the state which remains maximally entangled regardless of the choice of mode.
	Interestingly, the resulting state is not defined by uniformly distributed Schmidt coefficients, which would be the case for Bell-type states which are often considered to be maximally entangled.
	A witnessing approach further enables us to verify the MIQE characteristics and allows us to impose tight bounds to perturbations which may diminish our specific quantum characteristics.
	In Sec. \ref{sec:Generalization}, we generalize our approach to multiphoton and multimode states.
	From this generalization, we are then able to describe classes of states which are even invariant under nonunitary mode transformations, and which remain entangled under arbitrary partitioning of subsystems that constitute various forms of entanglement, such as partial and full inseparability.
	Based on our analysis for pure states, we end our discussions with a rigorous definition of mode-independent separability for mixed quantum states in Sec. \ref{sec:Mixed}, which supersedes the commonly applied definition of separability in a fixed modal reference frame.
	In our concluding discussion, Sec. \ref{sec:Conclusion}, we also outline potential generalizations to continuous-variable states and nonlinear mode transformations.
	Therefore, we develop a broadly applicable framework to define, generate, and verify MIQE of light.

%%%%%%%%%%%%%%%%%%%%%%%%%%%%%%%%%%%%%%%%%%%%%%%%%%%%%%%%
% Preliminaries
%%%%%%%%%%%%%%%%%%%%%%%%%%%%%%%%%%%%%%%%%%%%%%%%%%%%%%%%
\section{Mode dependence of entanglement}\label{sec:Preliminaries}

\subsection{Separability with respect to modes}

	The accepted definition of entanglement is based on rejecting the corresponding notion of classical correlation for separable states \cite{W89}.
	It is worth emphasizing that separability is a property of a state, not a system.
	This means that it presupposes a quantum system and a given set of degrees of freedom.
	With respect to the latter degrees of freedom, a separation of the composite system into subsystems is possible that ultimately defines the notion of a separable state \cite{W89}.
	For simplicity, we restrict ourselves to bipartite systems, the parts of which are labeled as $1$ and $2$, for the time being.
	In optical systems, the degrees of freedom are represented through well-defined, orthogonal optical modes.

	The definition of a separable state $\hat \sigma$ is given in terms of the possibility to write this state as a statistical mixture of pure product states, $|\psi^{(1)},\psi^{(2)}\rangle$, as
	\begin{align}
		\label{eq:DefSep}
		\hat\sigma=\int dP(\psi^{(1)},\psi^{(2)})\,|\psi^{(1)},\psi^{(2)}\rangle\langle\psi^{(1)},\psi^{(2)}|,
	\end{align}
	where $P$ represents a classical probability distribution \cite{W89}.
	Conversely, for an inseparable, likewise entangled, state, $\hat\rho$, such a convex decomposition does not exist.
	Rather, $P$ takes the form of a distribution which includes negativities \cite{STV98,SV09prime}; see Ref. \cite{SMBBS19} for a recent experimental reconstruction of such so-called entanglement quasiprobabilities.
	We also emphasize that a pure state $|\psi\rangle$ is entangled when $|\psi\rangle\neq |\psi^{(1)},\psi^{(2)}\rangle$ holds true for any $|\psi^{(1)}\rangle$ and $|\psi^{(2)}\rangle$.

	The optical modes we are considering are represented through the annihilation operators $\hat a_1$ and $\hat a_2$.
	A transformation between different, orthogonal mode decompositions is described through a unitary map,
	\begin{align}
		\begin{pmatrix}
			\hat b_1\\\hat b_2
		\end{pmatrix}
		=
		U
		\begin{pmatrix}
			\hat a_1\\\hat a_2
		\end{pmatrix},
	\end{align}
	where the annihilation operators of the resulting modes are identified through $\hat b_1$ and $\hat b_2$, and $U$ is the unitary transformation matrix, $U^\dag U=\mathrm{id}$.
	A convenient representation of the latter matrix for two modes is formulated in terms of transmission and reflection amplitudes, $t$ and $r$, respectively, with $|t|^2+|r|^2=1$.
	For instance, we can then write
	\begin{align}
		\label{eq:Transformation}
		t \hat a_1^\dag + r\hat a_2^\dag=\hat b_1^\dag
		\text{ and }
		t^\ast \hat a_2^\dag - r^\ast\hat a_1^\dag=\hat b_2^\dag
	\end{align}
	for the input-output relation of creation operators.

	Since such a mode transformation acts globally on multiple modes, it constitutes a nonseparable operation \cite{R97,VP98}.
	Consequently, the definition of separability also changes in this new reference frame.
	For instance, we have pure separable states which are given by $|\psi\rangle=|\psi^{(1)}_U,\psi^{(2)}_U\rangle$, where the index $U$ indicates that we operate in transformed degrees of freedom, i.e., a rotated mode basis.
	This notation is used throughout this paper.
	The ultimate aim of this paper is to construct states which are inseparable for any mode decomposition.

	As we are considering different mode decompositions, easily distinguishable notations are needed.
	For this reason, initial computational basis modes are labeled with ``$a$,'' unitary-transformed basis modes are labeled with ``$b$,'' and excited modes to generate photons are labeled with ``$c$.''
	The latter concept is studied in the following.

\subsection{Failures}

	Before we construct the desired family of states, we may consider specific examples and analyze why they cannot satisfy our demands.
	Obviously, the vacuum state $|\mathrm{vac}\rangle=|0,0\rangle=|0_U,0_U\rangle$, having a total photon number of zero, is factorizable in any mode basis.
	That is, the vacuum state does not have a preferred basis and exhibits no quantum correlations.

	The next example is a pure single-photon state, which has the general form
	\begin{align}
		\label{eq:1Phot}
		\lambda_{1,0}|1,0\rangle+\lambda_{0,1}|0,1\rangle
		=\left(\lambda_{1,0}\hat a_1^\dag+\lambda_{0,1}\hat a_2^\dag\right)|\mathrm{vac}\rangle,
	\end{align}
	which is entangled for nonzero Schmidt coefficients $\lambda_{1,0}$ and $\lambda_{0,1}$.
	Evidently, when choosing $t=\lambda_{1,0}$ and $r=\lambda_{0,1}$ for the relations in Eq. \eqref{eq:Transformation}, we can write the same state in the separable form, $\hat b_1^\dag|\mathrm{vac}\rangle=|1_U,0_U\rangle$.
	This representation confirms the known observation that single-photon states can be transformed into separable states via an appropriate mode transformation, as well as the other way around; see, e.g., Ref. \cite{TWC91}.
	From the quantum-field-theoretic perspective, this transformation yields the superposition mode $\hat c_1$ that is excited once to result in the single-photon state under study, i.e., applying $\hat c_1^\dag=\lambda_{1,0}\hat a_1^\dag+\lambda_{0,1}\hat a_2^\dag$ to vacuum, Eq. \eqref{eq:1Phot}.
	In conclusion, we can say at least two photons are required to achieve our goal.

	Nevertheless, even in the case of more than one photon, success is not guaranteed.
	Specifically, two forms of separability can occur in the scenario of two photons.
	First, we can have both photons in the same mode, $\hat c_1^\dag=\hat c_2^\dag=t\hat a_1^\dag+r\hat a_2^\dag$, which yields
	\begin{align}
		\label{eq:02PhotSep}
		\frac{\hat c_1^{\dag}\hat c_2^\dag}{\sqrt{2}}|\mathrm{vac}\rangle=t^2|2,0\rangle+\sqrt 2 t r |1,1\rangle+r^2|0,2\rangle
		=|2_U,0_U\rangle.
	\end{align}
	When $U$ is chosen such that it rotates $\hat b_1=\hat c_1$, we see that this state is factorizable.
	The second option is that we have one photon in each orthogonal mode, where $\hat c_1^\dag$ is chosen as before and $\hat c^\dag_2=-r^\ast\hat a_1^\dag+t^\ast\hat a_2^\dag$.
	This results in states of the form
	\begin{align}
		\nonumber
		\hat c_1^\dag\hat c_2^\dag|\mathrm{vac}\rangle
		=&-\sqrt{2}t r^\ast|2,0\rangle
		+\sqrt{2}t^\ast r|0,2\rangle+(|t|^2-|r|^2)|1,1\rangle
		\\\label{eq:11PhotSep}
		=&|1_U,1_U\rangle.
	\end{align}
	Again, this state is separable under the unitary mode transformation $U$.

	For example, a symmetric splitting of the modes used in Eq. \eqref{eq:11PhotSep} results in a so-called NOON state \cite{BKABWD00}, i.e., $(|2,0\rangle+e^{i\varphi}|0,2\rangle)/\sqrt 2$, for a local phase $\varphi$, and ignoring a global phase.
	This state is entangled in the computational basis, i.e., with respect to the basis modes $a$.
	It also keeps its form under some transformation, such as the eigenbasis of the Hamiltonian that corresponds to the beam splitter processes to generate this state.
	In particular, for any $|t|=|r|$, the above equation preserves the NOON-type structure.
	Yet, this state, which is obtained in the Hong-Ou-Mandel interference \cite{HOM87}, is separable as it is a result of two individual photons $|1_U,1_U\rangle$ which are combined on a general beam splitter represented by $U$, which is formally represented through Eq. \eqref{eq:11PhotSep}.
	Thus, this specific NOON state is separable for some mode bases.

	Therefore, we make the observation that a state the entanglement of which survives any classical mode transformation must not be an element of one of the two families of separable two-photon states in Eqs. \eqref{eq:02PhotSep} and \eqref{eq:11PhotSep}.

%%%%%%%%%%%%%%%%%%%%%%%%%%%%%%%%%%%%%%%%%%%%%%%%%%%%%%%%
% Two-photon, two-mode states
%%%%%%%%%%%%%%%%%%%%%%%%%%%%%%%%%%%%%%%%%%%%%%%%%%%%%%%%
\section{Mode-independent inseparability}\label{sec:MN2}

\subsection{Exciting nontrivially related optical modes}

	Because the above examples do not lead to states with the desired MIQE properties, we have to focus on two-photon states that excite optical modes which are nonparallel and nonorthogonal.
	Thus, we propose states of the form
	\begin{align}
		\label{eq:2PhotInsep}
	\begin{aligned}
		|\Psi\rangle
		=\mathcal N\hat c_1^\dag\hat c_2^\dag|\mathrm{vac}\rangle
		=\frac{\sqrt 2|2,0\rangle+\lambda|1,1\rangle}{\sqrt{2+|\lambda|^2}},
		\\
		\text{with }
		\hat c_1^\dag\propto\hat a_1^\dag
		\text{ and }
		\hat c_2^\dag\propto\hat a_1^\dag+\lambda\hat a_2^\dag,
	\end{aligned}
	\end{align}
	$\mathcal N=1/\sqrt{2+|\lambda|^2}$ being a normalization constant such that $\langle\Psi|\Psi\rangle=1$ holds true, and $\lambda\neq0$ defining an arbitrary complex number.
	This state  can be physically described as the state of a two-mode light field in which two photons live in nonparallel and nonorthogonal modes, $\hat c_1$ and $\hat c_2$ for $\lambda\neq0$.

	This simple but effective mechanism of exciting photons in nonparallel and nonorthogonal modes is the key concept that enables us to construct states with MIQE.
	This can be seen from the fact that any unitary transformations are going to preserve this structure of the involved modes $\hat c_1$ and $\hat c_2$.
	Thus, as we rigorously demonstrate in the following, no modal separation of the photons is possible.

\subsection{Witnessing mode-independent entanglement}

	For identifying entanglement in experiments, a convenient approach can be formulated in terms of measurable entanglement tests \cite{HHH96,T05,SV09}.
	Specifically, we employ a method related to entanglement witnesses in which bounds for separable states are computed and that can be violated with entangled states.
	Such entanglement criteria for a state $\hat\rho$ read
	\begin{align}
		\label{eq:EntanglementCriterion}
		\mathrm{tr}(\hat\rho\hat L)=\langle \hat L\rangle>g,
	\end{align}
	where $\hat L$ represents any observable under study and $g$ denotes the maximal expectation value for separable states \cite{T05,SV09},
	\begin{align}
		\label{eq:MaxSEVal}
		g=\sup_{\hat\sigma\text{ separable}}\mathrm{tr}(\hat\sigma\hat L).
	\end{align}
	Note that, because of convexity, it is sufficient to maximize over pure separable states, $\hat\sigma=|\psi^{(1)},\psi^{(2)}\rangle\langle \psi^{(1)},\psi^{(2)}|$.
	Moreover, the optimal expectation value $g$ can be obtained by solving the so-called separability eigenvalue equations \cite{SV09}, similarly to obtaining the maximal expectations values for arbitrary states through eigenvalue equations.

	Furthermore, for pure states, it can be shown that a viable test operator corresponds to the projector which maps onto the subspace spanned by the state $|\Psi\rangle$ itself \cite{SV09}, i.e., $\hat L=|\Psi\rangle\langle\Psi|$.
	As $|\Psi\rangle$ is normalized, we have $\langle\hat L\rangle=\mathrm{tr}(|\Psi\rangle\langle\Psi|\Psi\rangle\langle\Psi|)=1$.
	The bound for separable states $g$ in Eq. \eqref{eq:MaxSEVal} now provides a necessary and sufficient entanglement test because $g=1$ if and only if $|\Psi\rangle$ is separable \cite{SV09}, i.e., $|\langle\Psi|\psi^{(1)},\psi^{(2)}\rangle|^2=1$ when $|\Psi\rangle=|\psi^{(1)},\psi^{(2)}\rangle$.
	This is a result from the fact that the maximal expectation value $g$ of the projector $\hat L$ is identical with the maximal overlap of $|\Psi\rangle$ with separable states [see Eq. \eqref{eq:MaxSEVal}].
	Thus, we have for our choice of entanglement test and pure states the entanglement condition
	\begin{align}
		\label{eq:SimpleCrit}
		1>g,
	\end{align}
	and separability is identified through $\langle \hat L\rangle=1=g$.
	In particular, $g$ is the modulus square of the maximal Schmidt coefficient of $|\Psi\rangle$ \cite{SV09}.
	It is worth mentioning that any operator $\hat L$ for which the eigenspace corresponding to the maximal eigenvalue contains the vector $|\Psi\rangle$, but no factorizable (i.e., separable) vector, works as well as the test operator $\hat L=|\Psi\rangle\langle\Psi|$ selected here.

	We now apply our criteria to our family of states under study in a transformed modal basis, which also impacts the Schmidt decomposition.
	Inverting the transformation in Eq. \eqref{eq:Transformation}, i.e., $\hat a_1^\dag=t^\ast \hat b_1^\dag+r\hat b_2^\dag$ and $\hat a_2^\dag=t\hat b_2^\dag-r^\ast \hat b_1^\dag$, we can write the state in Eq. \eqref{eq:2PhotInsep} as
	\begin{align}
		|\Psi\rangle=\lambda_{2,0}|2_U,0_U\rangle+\lambda_{1,1}|1_U,1_U\rangle+\lambda_{0,2}|0_U,2_U\rangle,
	\end{align}
	which is already in the Schmidt decomposition of photon-number states $|n_U,2_U-n_U\rangle$ of the transformed modes, where
	\begin{align}
	\label{eq:SchmidtCoeff}
	\begin{aligned}
		\lambda_{2,0}=\frac{\sqrt 2t^\ast(t^\ast-\lambda r^\ast)}{\sqrt{2+|\lambda|^2}},\,
		\lambda_{0,2}=\frac{\sqrt 2r(r+\lambda t)}{\sqrt{2+|\lambda|^2}},
		\\\text{and}\quad
		\lambda_{1,1}=\frac{2t^\ast r+\lambda(|t|^2-|r|^2)}{\sqrt{2+|\lambda|^2}}.
	\end{aligned}
	\end{align}
	Indeed, we then get for the bound $g_U$ in the rotated basis
	\begin{align}
		\label{eq:PureStateBound}
		g_U=\max\{|\lambda_{2,0}|^2,|\lambda_{1,1}|^2,|\lambda_{0,2}|^2\}<1,
	\end{align}
	for any choice of $\lambda\neq0$; we refer to Appendix \ref{app:M2N2opt} for technical details.
	From the optimization over $U$ performed there, we also find that the maximal value for any choice of mode coefficients $t$ and $r$, i.e., unitary maps $U$, reads
	\begin{align}
		\label{eq:Bound2Phot}
		g_\mathrm{MI}=\max_{U} g_U=\max\left\{
			\frac{1}{2}+\frac{\sqrt{1{+}|\lambda|^2}}{2{+}|\lambda|^2},
			1-\frac{1}{2{+}|\lambda|^2}
		\right\}.
	\end{align}

	This mode-independent bound $g_\mathrm{MI}$ now represents the maximal expectation value of $\hat L$ for separable states with arbitrary mode decompositions.
	Again, we can readily confirm via Eq. \eqref{eq:Bound2Phot} that $g_\mathrm{MI}<1$ holds true for any $|\lambda|\neq0$; see also the following application in Sec. \ref{sec:OptCaseExample}.

	More generally, we can conclude for generally mixed states $\hat\rho$ that, as long as
	\begin{align}
		\label{eq:MixedState}
		\langle\hat L\rangle=\langle\Psi|\hat\rho|\Psi\rangle>g_\mathrm{MI}
	\end{align}
	holds true, we have certified MIQE of the mixed state $\hat\rho$.
	The above inequality can be equally understood in terms of a minimal fidelity $g_\mathrm{MI}$ of the state $\hat\rho$ and the target state $|\Psi\rangle$ which has to be exceeded to ensure MIQE.
	Note that it is sufficient to achieve this fidelity with any of the (transformed) states in the considered class of pure two photons.
	We also emphasize that this fidelity also provides a bound to imperfections which can be tolerated in a realistic implementation.

% note: wrong placement %%%%%%%%%%%%%%%%%%%%%%%%%%%
\begin{figure*}
	\includegraphics[width=\textwidth]{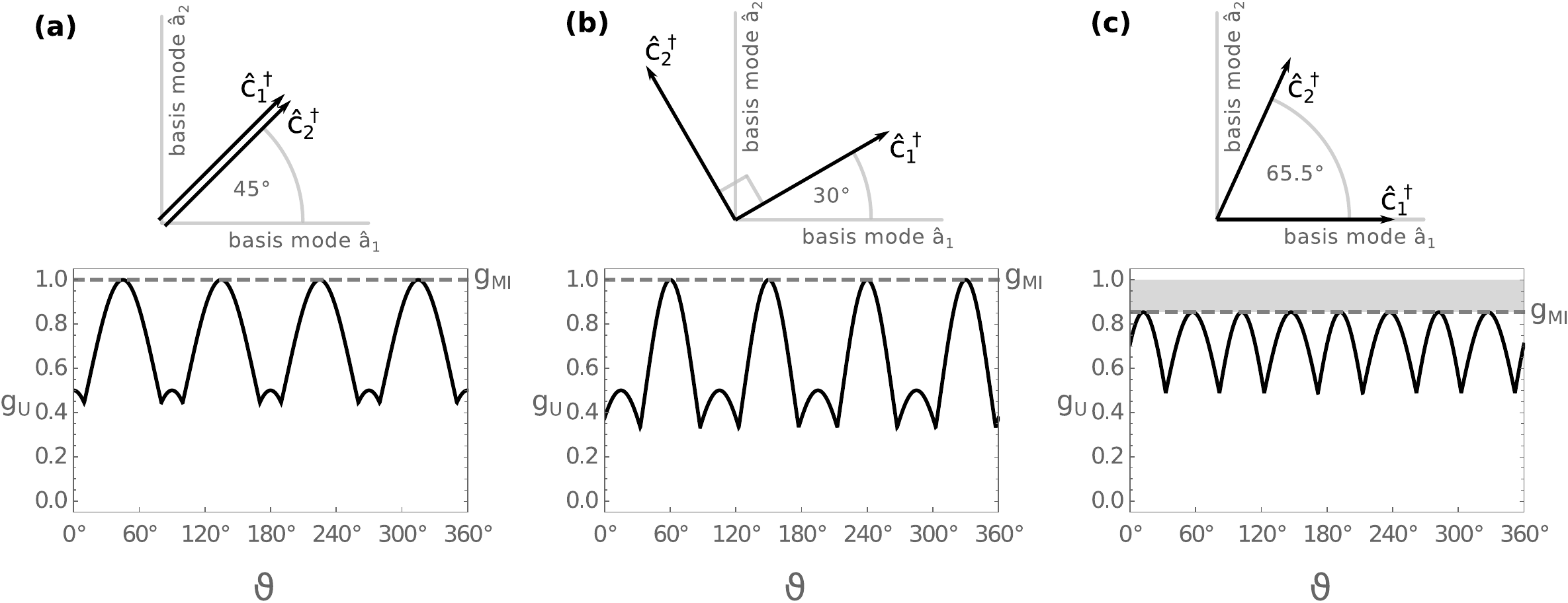}
	\caption{
		MIQE test, $g_\mathrm{MI}<1$.
		The top row depicts the modes in which each of the two photons is excited, where arrows indicate $\hat c^\dag_1$ and $\hat c^\dag_2$ and the basis modes are $\hat a_1$ and $\hat a_2$, respectively.
		The entanglement test $g_U<1$ in the bottom row is performed with respect to the unitary mode decomposition $\hat b_1=\cos\vartheta\,\hat a_1+\sin\vartheta\,\hat a_2$ and $\hat b_2=-\sin\vartheta\,\hat a_1+\cos\vartheta\,\hat a_2$.
		Column (a) corresponds to photons which are excited along the same mode [see Eq. \eqref{eq:02PhotSep} for a $45^\circ$ rotation].
		Column (b) describes photons which are excited in perpendicular modes [see Eq. \eqref{eq:11PhotSep} for a $30^\circ$ rotation].
		Column (c) shows photons which are excited in modes with an angle of $65.5^\circ$ between them, constituting the optimal mode-independent scenario, cf. Eqs. \eqref{eq:2PhotInsep} and \eqref{eq:lambdaOpt}.
		While being entangled ($g_U<1$) for most angles in the cases (a) and (b), the states are indeed separable ($g_U=1$) for some angles $\vartheta$ (e.g., $\vartheta=45^\circ$ and $\vartheta=60^\circ$), resulting in $g_\mathrm{MI}=1$ (dashed lines).
		By contrast, we have a finite offset (gray area) for the scenario (c), with $g_U\leq g_\mathrm{MI}<1$ [see also Eq. \eqref{eq:gOpt}], proving the state's inseparability regardless of the mode decomposition.
	}\label{fig:fig}
\end{figure*}

\subsection{Optimal mode-independent entanglement}\label{sec:OptCaseExample}

	In Appendix \ref{app:M2N2opt}, we further determined the parameter $\lambda$ for which this bound $g_\mathrm{MI}$ can be beaten with the largest difference to one, the separable bound [see Eq. \eqref{eq:SimpleCrit}].
	We find that the value
	\begin{align}
		\label{eq:lambdaOpt}
		|\lambda|=\sqrt{2\left(1+\sqrt2\right)}\approx2.197
	\end{align}
	is optimal, where the phase $\arg(\lambda)$ can be chosen arbitrarily.
	This then gives a bound, i.e., minimal fidelity with the target state $|\Psi\rangle$, as
	\begin{align}
		\label{eq:gOpt}
		g_\mathrm{MI}=\frac{2+\sqrt 2}{4}\approx85.4\%.
	\end{align}
	This certainly represents a comparably challenging bound, yet not an infeasible one.

	For example, one way to construct such a state is by using photon-addition protocols \cite{K08} and polarization degrees of freedom, e.g., $\hat a_1$ and $\hat a_2$ corresponding to vertical and horizontal polarization.
	In the first step, one adds a photon in horizontal polarization, approximating the $c_1^\dag=\hat a_1^\dag$ operation.
	Applying wave plates, the polarization is then rotated such that the horizontal component is mapped onto $(\hat a_1+\lambda^\ast\hat a_2)/\sqrt{1+|\lambda|^2}$.
	Note that the optimal value of $|\lambda|$ given above corresponds to a rotation of ca. $65.5^\circ$.
	Finally, performing another photon addition realizes the excitation of a photon along this new horizontal polarization, $\hat c_2^\dag\propto\hat a_1^\dag+\lambda\hat a_2^\dag$.
	Ideally, this produces the state as defined in Eq. \eqref{eq:2PhotInsep}.
	As long as all imperfections one encounters in this state preparation are below the threshold set by the fidelity $85.4\%$, a slightly different state is generated which, however, exhibits the same desired MIQE [see Eq. \eqref{eq:MixedState}].
	See Ref. \cite{SVA14} for an in-depth analysis of realistic photon-addition processes.

	In Fig. \ref{fig:fig}, we compare different types of two photon states to depict the results of our analysis.
	When the two photons are excited in the same mode [see Fig. \ref{fig:fig}(a), where $\hat c_1=(\hat a_1+\hat a_2)/\sqrt 2=\hat c_2$], separability can be observed for specific mode decompositions;
	when the two photons are excited in orthogonal modes [see Fig. \ref{fig:fig}(b), where $\hat c_1=(\sqrt{3}\hat a_1+\hat a_2)/2$ and $\hat c_2=(-\hat a_1+\sqrt{3}\hat a_2)/2$], separability also occurs for other specific mode decompositions.
	In both scenarios, this can be concluded from the fact that the fidelity of the state under study with separable states becomes $g_U=1$ for specific angles $\vartheta$ between the modes $\hat b_1$ and $\hat b_2$.
	The latter modes define the transformed degrees of freedom which, as we want to stress, also change the definition of separability, $|\Psi\rangle=|\psi_U^{(1)},\psi_U^{(2)}\rangle$, to this rotated system.
	However, when the photons are excited along fields which are nonparallel and nonorthogonal, MIQE is detected, $g_U\leq g_\mathrm{MI}<1$.
	This scenario is shown in Fig. \ref{fig:fig}(c) for the optimal choice of $\lambda$.

	Let us make some extra remarks on Fig. \ref{fig:fig} and the two-mode, two-photon case.
	Because of maximal and minimal values occurring only for $\mathrm{Im}[\lambda t r^\ast]=0$ (see Appendix \ref{app:M2N2opt}), we set without a loss of generality $t=\cos\vartheta$  and $r=\sin\vartheta$ for our choice $\lambda\approx2.197$.
	It is also worth emphasizing that the optimal choice of $\lambda$ yields the state $|\Psi\rangle\propto|2,0\rangle+[1+\sqrt{2}]|1,1\rangle$, which has nonuniformly distributed Schmidt coefficients and is therefore not a Bell-type state.
	Furthermore it is worth mentioning that, for $|\Psi\rangle=(2+|\lambda|^2)^{-1/2}(\sqrt{2}|2,0\rangle+\lambda|1,1\rangle)$ in the limits $|\lambda|\to 0$ and $|\lambda|\to\infty$, we retrieve the parallel and orthogonal two-photon cases, respectively.

%%%%%%%%%%%%%%%%%%%%%%%%%%%%%%%%%%%%%%%%%%%%%%%%%%%%%%%%
% Generalization
%%%%%%%%%%%%%%%%%%%%%%%%%%%%%%%%%%%%%%%%%%%%%%%%%%%%%%%%

\section{Multipartite multiphoton states}\label{sec:Generalization}

\subsection{Generalization}

	Based on our previous considerations, we can now generalize the procedure to obtain states with MIQE.
	For doing so, we consider $N$ photon states in $M$ modes, described through $\hat a_l$ for $l\in\{1,\ldots,M\}$.
	Specifically, we excite the $k$th photon ($k\in\{1,\ldots,N\}$) according to
	\begin{align}
		\label{eq:MNmode}
		\hat c_k^\dag=\sum_{l=1}^M \Gamma_{k,l}\hat a_l^\dag,
	\end{align}
	where vectors $(\Gamma_{k,l})_{l\in\{1,\ldots,M\}}$ that describe the excited field modes are linearly independent but not orthogonal.
	This results in the construction of the state
	\begin{align}
		\label{eq:MNstate}
		|\Psi_{M,N}\rangle=\mathcal N \hat c_1^\dag\cdots\hat c_N^\dag |\mathrm{vac}\rangle.
	\end{align}

	In principle, we can then proceed as discussed for the case of two photons in two modes.
	In particular, the entanglement analysis with respect to unitary transformations of modes, $\hat a_l^\dag=\sum_{j=1}^M U_{l,j}\hat b_j^\dag$, can be performed and optimized.
	For this purpose, it is helpful to recall that the generated states in the rotated basis always exhibit a decomposition with the transformed photon-number states $|n_U^{(1)},\ldots,n_U^{(M)}\rangle$ restricted to $N=n_U^{(1)}+\cdots+n_U^{(M)}$, which is convenient for constructing entanglement witnesses as done before but for a multipartite system \cite{SV13,GVS18}.
	In the following, we discuss a few interesting examples in more detail;
	see Appendix \ref{app:Algebra} for a full algebraic characterization of the general case.

\subsection{Examples}

\subsubsection{Three photons in two modes}

	As a first example, we study a state which goes beyond two photons but still remains in the two-mode case.
	Thus, we set $M=2$ and $N=3$ and consider
	\begin{align}
	\begin{aligned}
		|\Psi_{2,3}\rangle
		=&\mathcal N\hat a_1^\dag(\hat a_1^\dag+\hat a_2^\dag)(\hat a_1^\dag+i\hat a_2^\dag)|\mathrm{vac}\rangle
		\\
		=&\frac{\sqrt{3}|3,0\rangle+(1+i)|2,1\rangle+i|1,2\rangle}{\sqrt 6}.
	\end{aligned}
	\end{align}
	Similar to the two-photon case, we can observe that this state exhibits entanglement regardless of the choice of unitary mode transformations because we have pairwise nonparallel and nonorthogonal excitation, $\hat c_1^\dag=\hat a_1^\dag$ and $\hat c_2^\dag=\hat a_1^\dag+\hat a_2^\dag$, as well as the additional component $\hat c_3^\dag=\hat a_1^\dag+i\hat a_2^\dag$.

	In addition to this, we can even exclude general linear transformations [i.e., elements of the general linear group $\mathrm{GL}(M)$] for this state, which exceed the set of unitary transformations [i.e., $\mathrm{U}(M)\subsetneq \mathrm{GL}(M)$].
	This can be seen from the fact that, even if we consider $\hat c_1$ and $\hat c_2$ as a nonorthonormal mode basis, we still have to express the third mode as a superposition of the former ones, $\hat c_3=(1+i)\hat c_1-i\hat c_2$.
	This results in $|\Psi_{2,3}\rangle\neq|2_U,1_U\rangle$ and $|\Psi_{2,3}\rangle\neq|3_U,0_U\rangle$ for any unitary and general linear transformation $U$.
	Thus, the three-photon state $|\Psi_{2,3}\rangle$ is even invariant under general linear transformations.

\subsubsection{Two photons in three modes}

	Now, we may focus on more than two modes.
	For example, $N=2$ photons could be distributed over $M=3$ modes, such as
	\begin{align}
	\begin{aligned}
		|\Psi_{3,2}\rangle
		=&\mathcal N\hat a_1^\dag(\hat a_1^\dag+\hat a_2^\dag+\hat a_3^\dag)|\mathrm{vac}\rangle
		\\
		=&\frac{\sqrt 2|2,0,0\rangle+|1,1,0\rangle+|1,0,1\rangle}{2}.
	\end{aligned}
	\end{align}
	This state can be used to demonstrate properties which are genuine to multimode entanglement.

	For this purpose, we consider the unitary basis transformation $\hat b_1=\hat a_1$, $\hat b_2=(\hat a_2+\hat a_3)/\sqrt 2$, and $\hat b_3=(-\hat a_2+\hat a_3)/\sqrt 2$.
	Then, we get
	\begin{align}
		|\Psi_{3,2}\rangle=\frac{|2_U,0_U,0_U\rangle+|1_U,1_U,0_U\rangle}{\sqrt 2}.
	\end{align}
	In this basis, we can see that the state of the third system factorizes, $|\Psi\rangle=|\Psi^{(1,2)},\psi^{(3)}_U\rangle$ with $|\psi^{(3)}_U\rangle=|0_U\rangle$, implying partial separability.
	The first and second modes do not separate as we have a state as given in Eq. \eqref{eq:2PhotInsep} for $\lambda=\sqrt2$, hence $|\Psi^{(1,2)}\rangle\neq|\psi_U^{(1)},\psi_U^{(2)}\rangle$ for any $U$.
	Therefore, this example exhibits MIQE with respect to full separability, but it is partially separable in a mode-independent manner at the same time.

\subsubsection{Three photons in three modes}

	Consequently, we may consider three excitations in three modes, $M=N=3$.
	For instance, we can have a state like
	\begin{align}
	\begin{aligned}
		&|\Psi_{3,3}\rangle
		=\mathcal N\hat a_1^\dag(\hat a_1^\dag+\hat a_2^\dag)(\hat a_1^\dag+\hat a_2^\dag-\hat a_3^\dag)|\mathrm{vac}\rangle
		\\
		=&\frac{1}{\sqrt{19}}(
			\sqrt{6}|3,0,0\rangle
			+2\sqrt{2}|2,1,0\rangle
		\\ &\phantom{\sqrt{19}}
			-\sqrt{2}|2,0,1\rangle
			+\sqrt{2}|1,2,0\rangle
			-|1,1,1\rangle
		).
	\end{aligned}
	\end{align}
	As no unitary mode decomposition can separate this state, not even partially, because of the nonparallel and nonorthogonal nature of all three generated photons, this state is entangled with respect to mode-independent full and partial separability.

	More generally, we require $M\leq N$ for achieving MIQE with respect to any mode partitioning.
	See Appendix \ref{app:Algebra} in this context and for the general characterization of $|\Psi_{M,N}\rangle$ using the $QR$ decomposition, being more suitable for MIQE than the Schmidt (likewise, singular value) decomposition.
	Furthermore, a suitable test operator for these multimode states is $\hat L=|\Psi_{M,N}\rangle\langle\Psi_{M,N}|$ for which the bounds for partial and full separability can be obtained in a similar fashion as demonstrated for the bipartite case \cite{SV13,GVS18}.
	Thus, the general approach, given by the excitation of photons along linearly independent and nonorthogonal modes as described in Eq. \eqref{eq:MNmode} and resulting in the state in Eq. \eqref{eq:MNstate}, extends the concept of MIQE straightforwardly to the multimode and multiphoton scenario.

%%%%%%%%%%%%%%%%%%%%%%%%%%%%%%%%%%%%%%%%%%%%%%%%%%%%%%%%
% Mixed
%%%%%%%%%%%%%%%%%%%%%%%%%%%%%%%%%%%%%%%%%%%%%%%%%%%%%%%%
\section{Redefining separability for mode independence}\label{sec:Mixed}

	So far, we mainly focused on pure states in our analysis.
	But the definition of a separable state includes mixed ones [Eq. \eqref{eq:DefSep}] that are described as statistical mixtures of pure product states \cite{W89}.
	In addition to this, we allowed for unitary transformation of modes to discern states that are entangled regardless of the choice of modal decomposition from those which are separable in at least one mode basis.

	To reflect the latter feature, we may propose an alternative approach for defining separability of states which is mode-independent.
	Namely, we say a state $\hat\sigma_\mathrm{MI}$ exhibits mode-independent separability if it can be represented as a statistical mixture of pure states which are separable under mode transformations.
	Therefore, we may describe such states in the two-mode scenario as
	\begin{align}
		\label{eq:MISepDef}
		\hat\sigma_\mathrm{MI}=\int dP(\psi^{(1)}_U,\psi^{(2)}_U,U)|\psi^{(1)}_U,\psi^{(2)}_U\rangle\langle \psi^{(1)}_U,\psi^{(2)}_U|,
	\end{align}
	where $P$ is a probability distribution over not only product states but also unitary mode transformations $U$.

	In comparison to separability for a fixed separation of modes in Eq. \eqref{eq:DefSep}, the revised notion of mode-independent separability in Eq. \eqref{eq:MISepDef} is based on states for which it is sufficient that they are separable in one mode decomposition.
	This also means that a state $\hat\rho_\mathrm{MI}$ can be entangled with respect to some mode bases.
	For example, a mixture of the states with different total photon numbers, each being separable in a different mode basis, is separable with respect to definition \eqref{eq:MISepDef} but entangled for each basis choice, following, for example, from the partial transposition criterion \cite{P96}, which is a special case of the approach in Ref. \cite{SV09}.
	Finally, definition \eqref{eq:MISepDef} straightforwardly generalizes to arbitrary multimode scenarios and is not restricted to any fixed photon numbers.

%%%%%%%%%%%%%%%%%%%%%%%%%%%%%%%%%%%%%%%%%%%%%%%%%%%%%%%%
% Conclusions
%%%%%%%%%%%%%%%%%%%%%%%%%%%%%%%%%%%%%%%%%%%%%%%%%%%%%%%%
\section{Summary and conclusion}\label{sec:Conclusion}

	We devised a technique to construct photonic quantum states which exhibit inseparability that does not depend on the modal decomposition, where the individual modes define the separate parties.
	For this purpose, we considered multiphoton states in general multimode systems and analyzed how their entanglement transforms under unitary mode transformations.
	By combining photons which are, interestingly, defined over nonparallel and nonorthogonal modes, we were able to formulate states with the desired entanglement properties.

	Using a witnessing approach, we were further able to derive universal bounds for the considered families of states that verify entanglement for any choice of basis modes in terms of fidelities.
	For the specific, yet fundamental case of two photons distributed over two modes, we explicitly discussed the optimal case for getting the most robust state with MIQE.
	In particular, it was found that the two involved photons are excitations of modes with an angle of $65.5^\circ$ between them.
	We discussed potential realizations of this scenario by employing polarization degrees of freedom and photon-addition protocols.
	This certifies that one can, in principle, generate photonic states with entanglement properties which do not require any specific optical-mode basis.
	This is in contrast to the typical scenario in which entangled photonic states are considered that are, however, separable for some choices of modes \cite{TWC91}.

	Beyond the essential example of two photons distributed in two modes, we showed that our MIQE method is easily scalable to multiple photons and multimode light fields.
	For instance, we showed that one can build three-photon states in two modes, which are entangled when performing not only unitary mode transformation but arbitrary linear transformations of modes.
	Moreover, we analyzed how the number of photons in multimode systems affects the potential to generate different forms of partial and full entanglement.
	Specifically, we found that the number modes must not exceed the number of linearly independent, but nonparallel photons which need to be excited to generate a state which is entangled with respect to arbitrary forms of multipartite entanglement.

	Here, we deliberately choose to restrict ourselves to states defined over finite-dimensional subspaces \cite{SV09finite} with a fixed photon number for introducing the concept of MIQE, which was later completed with the general definition of mode-independent separability and inseparability.
	The latter definition is stronger than the typical applied definitions in the sense that it leads to quantum correlations which are decoupled from the representation of an optical state in a specific mode basis.
	Thus, MIQE represents a previously unknown flavor of global quantum correlations that is independent of mode superposition as allowed by classical optics.

	In the future, one might additionally analyze, for example, families of continuous-variable states in detail.
	The two-mode case suggests, however, that non-Gaussian states are required to obtain MIQE, similarly to the demanded resources for realizing universal quantum computation \cite{LB99}.
	This observation follows from techniques similar to the ones presented here \cite{SV13} to analyze separability for Gaussian states under mode transformations, which additionally shows how our method can be adapted to other scenarios beyond states probed here.
	This fact follows similarly from the Bloch-Messiah and Williamson decomposition; see Ref. \cite{WFPT17} in this context.
	In addition, one could extend the range of operations, e.g., to general symplectic operations (i.e., multimode squeezing), which are considered as classical \cite{SV15} to analyze even more robust forms of entanglement.
	It would be also interesting for upcoming studies to investigate how nonorthogonal and nonparallel excitations of quantum fields generalize to fermionic systems and other bosonic field theories as one would expect that similarly universal entanglement features as introduced here might appear in such systems.

	In conclusion, we established a method to construct and analyze states of light with MIQE.
	As outlined above, our approach has the potential to inspire future research and might be accessible with state-of-the-art experimental techniques.
	Furthermore, and because of the mode independence and scalability, the constructed families of multiphoton states can also lead to previously inconceivable quantum communication protocols which exploit this unique resource of MIQE.

%%%%%%%%%%%%%%%%%%%%%%%%%%%%%%%%%%%%%%%%%%%%%%%%%%%%%%%%
% Acknowledgments
%%%%%%%%%%%%%%%%%%%%%%%%%%%%%%%%%%%%%%%%%%%%%%%%%%%%%%%%
\begin{acknowledgments}
	The authors are grateful to Norbert L\"{u}tkenhaus and Benjamin Brecht for valuable comments and enlightening discussions.
	The Integrated Quantum Optics group acknowledges financial support from the Gottfried Wilhelm Leibniz-Preis (Grant No. SI1115/3-1) and the European Commission through the ERC project \mbox{QuPoPCoRN} (Grant No. 725366).
\end{acknowledgments}

\appendix

%%%%%%%%%%%%%%%%%%%%%%%%%%%%%%%%%%%%%%%%%%%%%%%%%%%%%%%%
% Optimal M=2,N=2 
%%%%%%%%%%%%%%%%%%%%%%%%%%%%%%%%%%%%%%%%%%%%%%%%%%%%%%%%
\section{Optimal two-photon, two-mode case}\label{app:M2N2opt}

	Based on the Schmidt coefficients in Eq. \eqref{eq:SchmidtCoeff}, we define three functions,
	\begin{subequations}
	\begin{align}
		\Lambda_{2,0}=&|\lambda_{2,0}|^2=\frac{
			2|t|^2(
				|t|^2{+}|\lambda|^2|r|^2{-}2\mathrm{Re}[\lambda t r^\ast]
			)
		}{2+|\lambda|^2},
		\\
		\Lambda_{2,0}=&|\lambda_{2,0}|^2=\frac{
			2|r|^2(
				|r|^2{+}|\lambda|^2|t|^2{+}2\mathrm{Re}[\lambda t r^\ast]
			)
		}{2+|\lambda|^2},
		\\\nonumber
		\Lambda_{1,1}=&|\lambda_{1,1}|^2=\frac{
			4|t|^2|r|^2
			{+}|\lambda|^2(|t|^2{-}|r|^2)^2
		}{2+|\lambda|^2}
		\\
		&\phantom{|\lambda_{1,1}|^2=}+\frac{
			4(|t|^2{-}|r|^2)\mathrm{Re}[\lambda t r^\ast]
		}{2+|\lambda|^2}.
	\end{align}
	\end{subequations}
	In the following, we optimize these expressions to find a $\lambda$ which minimizes $g_\mathrm{MI}$ in Eq. \eqref{eq:PureStateBound} for arbitrary $t$ and $r$.
	In this context, recall that $g_U$ itself is defined via a maximum of $\Lambda_{2,0}$, $\Lambda_{0,2}$, and $\Lambda_{1,1}$ \cite{SV09}.

	A first observation we can make is that the above functions jointly take optimal values if $\mathrm{Im}[\lambda t r^\ast]=0$ holds true.
	Thus, without loss of generality, we can suppose that $t$, $r$, and $\lambda$ are real numbers.
	Conversely, because of $|t|^2+|r|^2=1$, we set $t=\cos[\vartheta]$ and $r=\sin[\vartheta]$.
	Now, we can rewrite the functions under study as
	\begin{subequations}
	\begin{align}
		\Lambda_{2,0}=&\frac{1}{2}\frac{\left(
			1+\cos[2\vartheta]-\lambda\sin[2\vartheta]
		\right)^2}{2+\lambda^2},
		\\
		\Lambda_{0,2}=&\frac{1}{2}\frac{\left(
			1-\cos[2\vartheta]+\lambda\sin[2\vartheta]
		\right)^2}{2+\lambda^2},
		\\
		\Lambda_{1,1}=&\frac{\left(
			\sin[2\vartheta]+\lambda\cos[2\vartheta]
		\right)^2}{2+\lambda^2}.
	\end{align}
	\end{subequations}

	From the derivative of each function, and for the optimal cases (i.e., vanishing derivatives), we then get optimal arguments $\vartheta$ through the relations
	\begin{align}
		\text{either }
		\lambda=-\frac{\sin[2\vartheta]}{\cos[2\vartheta]}
		\text{ or }
		\lambda=\frac{\cos[2\vartheta]}{\sin[2\vartheta]}.
	\end{align}
	For these two cases, and after some algebra, we get $g_\mathrm{MI}$, i.e., the maximum of our functions, as
	\begin{align}
		g_\mathrm{MI}=\max\left\{
			\frac{1}{2}+\frac{\sqrt{\lambda^2+1}}{2+\lambda^2},
			1-\frac{1}{2+\lambda^2}
		\right\}.
	\end{align}
	We further want to know in which case $g_\mathrm{MI}$ is minimal for any mode transformation.
	As both values that determine $g_\mathrm{MI}$ have opposite monotonic behaviors and the minimum thereof is required, we consider the point at which both expressions are identical.
	For simplicity, we may introduce the substitution $x=\sqrt{1+\lambda^2}$, which is greater than one because of $\lambda\neq0$.
	This then gives two solutions, one of which does not satisfy $x>1$, leaving us with $x=1+\sqrt{2}$.

	From those considerations, we now obtain the optimal value of $\lambda$.
	Its amplitude reads
	\begin{align}
		|\lambda|=\sqrt{2\left(1+\sqrt 2\right)},
	\end{align}
	while the phase can be chosen arbitrarily.
	The resulting $\lambda$ then also defines the optimal state via Eq. \eqref{eq:2PhotInsep} and the mode-independent lower bound for separability, $g_\mathrm{MI}=(2+\sqrt 2)/4$.

%%%%%%%%%%%%%%%%%%%%%%%%%%%%%%%%%%%%%%%%%%%%%%%%%%%%%%%%
% General M,N 
%%%%%%%%%%%%%%%%%%%%%%%%%%%%%%%%%%%%%%%%%%%%%%%%%%%%%%%%
\section{General $M$-mode, $N$-photon scenario}\label{app:Algebra}

	Here, we characterize states of the form \eqref{eq:MNstate}, with excited modes as given in Eq. \eqref{eq:MNmode}.
	The characterization here is an independent algebraic approach, complementing the experimentally accessible witness-based method used previously.

\subsection{Separability via $\Gamma$}

	An $M$-mode state with exactly $N$ photons can be expanded via $|n_U^{(1)},\ldots,n_U^{(M)}\rangle$, where $n_U^{(1)}+\cdots+n_U^{(M)}=N$.
	Let us consider a bipartition, $|\Psi_{M,N}\rangle=|\psi_U^{(1,\ldots,\tilde M)},\psi_U^{(\tilde M+1,\ldots,M)}\rangle$.
	Since $|n_U^{(1)},\ldots,n_U^{(\tilde M)}\rangle$ for $n_U^{(1)}+\cdots+n_U^{(\tilde M)}=\tilde N$ and $|n_U^{(\tilde M+1)},\ldots,n_U^{(M)}\rangle$ for $n_U^{(\tilde M+1)}+\cdots+n_U^{(M)}=N-\tilde N$ are orthonormal in each subsystems for different $\tilde N$, the photon-number basis represents a Schmidt decomposition of the state.
	The state is thus biseparable (Schmidt rank one) when in each subsystem $\tilde N$ and $N-\tilde N$ are fixed.
	The same follows by induction for arbitrary multipartitions through further separations.

	Let us denote with $A$ and $B$ the two subsystems for convenience, and similarly $N_A=\tilde N$ and $N_B=N-\tilde N$ as well as $M_A=\tilde M$ and $M_B=M-\tilde M$.
	Since the number of photons in each part of a biseparable state $|\Psi_{M,N}\rangle=|\psi_U^{(A)},\psi_U^{(B)}\rangle$ has a fixed photon number in each part, the matrix $\Gamma$ then can be put in the general block form
	\begin{align}
		\label{eq:SepDelta}
		\Gamma=
		\begin{pmatrix}
			\Gamma_{A,A} & 0
			\\
			0 & \Gamma_{B,B}
		\end{pmatrix},
	\end{align}
	where $\Gamma_{j,j}\in\mathbb C^{N_j\times M_j}$ for $j\in\{A,B\}$.

\subsection{$QR$ decomposition}

	The $QR$ decomposition of a matrix $T\in\mathbb C^{M\times N}$ is an algebraic tool to analyze our states, represented through the rectangular matrix $\Gamma$.
	See, e.g., Ref. \cite{HJ13} for details on the $QR$ decomposition.
	The $QR$ decomposition theorem states there is a unitary $Q\in\mathbb C^{M\times M}$ and an upper triangular matrix $R\in\mathbb C^{N\times M}$ such that $T=QR$.
	For us, this means that we can identify $\Gamma=(\Gamma_{k,l})_{k\in\{1,\ldots,N\},l\in\{1,\ldots,M\}}=T^\dag$ [Eq. \eqref{eq:MNmode}], the mode transformation $U=Q^\dag$, and the lower triangular matrix $\Delta=R^\dag$, which gives the $QR$ decomposition as
	\begin{align}
		\Gamma=\Delta\, U.
	\end{align}

	Furthermore, it does not make a difference to exchange the excitations, $\hat c_{k}^\dag\hat c_{k'}^\dag=\hat c_{k'}^\dag\hat c_{k}^\dag$, which allows us to sort the rows of $\Gamma$ such that $\Gamma_k=(\Gamma_{k,l})_{l\in\{1,\ldots,M\}}$ for $k$ can be sorted in such a way that $\Gamma_k$ either is linearly dependent on $\Gamma_{k'}$ for $k'<k$ or represents a linearly independent vector to all previous ones.
	Note that potential elements with $\hat c_k=0$ (i.e., $\Gamma_k=0$ represents a row of zeros) can be discarded without a loss of generality.
	Using this sorting, the structure of $\Delta$ from the $QR$ decomposition takes a generalized lower triangular form,
	either
	\begin{align}
		\label{eq:Delta1}
		\Delta=&\Gamma U^\dag=
		\begin{pmatrix}
			\ast & 0 & \cdots
			\\
			\vdots & \vdots
			\\
			\ast & 0 & \cdots
			\\
			\ast & \ast & 0 & \cdots
			\\
			\vdots & \vdots & \vdots
			\\
			\ast & \ast & 0 & \cdots
			\\
			\ast & \ast & \ast & 0 & \cdots
			\\
			\vdots & \vdots & & \ddots & \ddots
			\\
			\ast & \ast & \cdots & \ast & \ast & 0 & \cdots
		\end{pmatrix}
	\end{align}
	or
	\begin{align}
		\label{eq:Delta2}
		\Delta=&\Gamma U^\dag=
		\begin{pmatrix}
			\ast & 0 & \cdots
			\\
			\vdots & \vdots
			\\
			\ast & 0 & \cdots
			\\
			\ast & \ast & 0 & \cdots
			\\
			\vdots & & \ddots & \ddots
			\\
			\ast & \ast & \cdots & \ast & 0
			\\
			\ast & \ast & \cdots & \ast & \ast
			\\
			\vdots & \vdots & & \vdots & \vdots
			\\
			\ast & \ast & \cdots & \ast & \ast
		\end{pmatrix},
	\end{align}
	where ``$\ast$'' represents an arbitrary entry.
	The last entry ``$\ast$'' in each row is nonzero; see the constructive proof in Ref. \cite{HJ13} for the $QR$ decomposition for details.
	In connection with separability, we can also easily observe that a $QR$ decomposition of the individual components $\Gamma_{A,A}$ and $\Gamma_{B,B}$ in Eq. \eqref{eq:SepDelta} is possible in the same manner.

\subsection{Conclusions}

	From the prior general analysis, we can deduce some special cases of particular interest.
	First, if there is a zero column [see Eq. \eqref{eq:Delta1}], the respective mode can be separated as a factor $|0_U\rangle$.
	In this case, we have at least partial separability.
	Such a zero column always exists for $M>N$.
	Second, if all $\Gamma_k$ are linearly independent, also meaning $M=N$, the size of each block of rows in Eqs. \eqref{eq:Delta1} and \eqref{eq:Delta2} is one.
	This also means that we have a strictly lower triangular (defined by nonzero diagonal entries) \cite{HJ13}.
	The same applies to the $QR$ decomposition of the individual blocks in the case that Eq. \eqref{eq:SepDelta} applies, further implying $M_j=N_j$ for $j\in\{A,B\}$.

	Finally, assume that we have performed a $QR$ decomposition for $N=M$ linearly independent mode vectors that are pairwise nonorthogonal, and we attempt to perform a bipartition into $A$ and $B$ via another unitary $V$.
	In this case, we may decompose $\Delta=\left(\begin{smallmatrix} \Delta_{A,A} & 0 \\ \Delta_{B,A} & \Delta_{B,B} \end{smallmatrix}\right)\in\mathbb C^{M\times M}$.
	Therein, $\Delta_{A,A}\in\mathbb C^{M_A\times M_A}$ and $\Delta_{B,B}\in\mathbb{C}^{M_B\times M_B}$ for some $M_A+M_B=M$ are strictly lower triangular because of linear independence, and $\Delta_{B,A}\neq0$ holds true because of the pairwise nonorthogonal nature.
	The unitary map to be constructed may be also decomposed in this block form $V=\left(\begin{smallmatrix} V_{A,A} & V_{B,A} \\ V_{B,A} & V_{B,B} \end{smallmatrix}\right)$.

	Then, if $\Delta\,V=\left(\begin{smallmatrix} \ast & 0 \\ 0 & \ast \end{smallmatrix}\right)$ would be true [representing a separable state; see Eq. \eqref{eq:SepDelta}], we would satisfy the conditions
	\begin{subequations}
	\begin{align}
		\label{eq:D1}
		0 =& \Delta_{A,A}V_{A,B},
		\\\label{eq:D2}
		0 =& \Delta_{B,A} V_{A,A}+\Delta_{B,B}V_{B,A}.
	\end{align}
	\end{subequations}
	In addition, the unitarity $V^\dag V=\left(\begin{smallmatrix} \mathrm{id} & 0 \\ 0 & \mathrm{id} \end{smallmatrix}\right)$ implies
	\begin{subequations}
	\begin{align}
		\label{eq:V1}
		\mathrm{id} =& V_{A,A}^\dag V_{A,A}+V_{B,A}^\dag V_{B,A},
		\\\label{eq:V2}
		\mathrm{id} =& V_{A,B}^\dag V_{A,B}+V_{B,B}^\dag V_{B,B},
		\\\label{eq:V3}
		0 =& V_{A,B}^\dag V_{A,A}+V_{B,B}^\dag V_{B,A}.
	\end{align}
	\end{subequations}
	Since $\Delta_{A,A}$ is a quadratic, strictly lower triangular matrix, the inverse exists and allows us to conclude from Eq. \eqref{eq:D1} the following:
	\begin{align}
	\begin{aligned}
		V_{A,B}=0
		\quad\stackrel{\text{Eq. \eqref{eq:V2}}}{\Longrightarrow}\quad
		V_{B,B}^{-1}=V_{B,B}^\dag
		\\
		\quad\stackrel{\text{Eq. \eqref{eq:V3}}}{\Longrightarrow}\quad
		V_{B,A}=0
		\quad\stackrel{\text{Eq. \eqref{eq:V1}}}{\Longrightarrow}\quad
		V_{A,A}^{-1}=V_{A,A}^\dag.
	\end{aligned}
	\end{align}
	Using the above relations, Eq. \eqref{eq:D2} then implies that $\Delta_{B,A}=0$ has to be true.
	This, however, contradicts the nonorthogonality requirement, $\Delta_{B,A}\neq0$.
	This proves that the desired unitary $V$ for the separation of $A$ and $B$ does not exist.

	Also note that the more general case $M>N$ [see Eq. \eqref{eq:Delta2}] follows equivalently.
	This can be easily shown by taking the $N$ rows which start a new block (i.e., are linearly independent from the previous rows) and considering the $N\times N$ submatrix in the same way as done for the case above, where $M=N$.

%%%%%%%%%%%%%%%%%%%%%%%%%%%%%%%%%%%%%%%%%%%%%%%%%%%%%%%%
% References
%%%%%%%%%%%%%%%%%%%%%%%%%%%%%%%%%%%%%%%%%%%%%%%%%%%%%%%%

\end{document}